\documentclass[11pt]{revtex4}
\usepackage{graphicx}
\newcommand{\be}{\begin{equation}}
\newcommand{\bea}{\begin{eqnarray}}
\newcommand{\ee}{\end{equation}}
\newcommand{\eea}{\end{eqnarray}}
\newcommand{\eps}{\epsilon}

\usepackage{amsmath}
\begin{document}

\title{Effective Hamiltonians for  fastly driven tight-binding chains}
\author{A.P. Itin$^{1,2}$\ and A.I. Neishtadt$^{2,3}$}
\affiliation{$^{1}$Zentrum f\"{u}r optische Quantentechnologien,
Universit\"{a}t Hamburg, Luruper Chaussee 149, 22761 Hamburg, Germany\\
$^2$Space Research Institute,  formerly Russian Academy of Sciences, Moscow, Russia\\
$^{3}$Department of Mathematical Sciences, Loughborough
University, Loughborough, LE11 3TU, UK.
}
\begin{abstract}

We consider a single particle tunnelling in a tight-binding model with nearest-neighbour couplings, in the presence of a periodic high-frequency force.
An effective Hamiltonian for the particle is derived using an averaging method resembling classical canonical perturbation theory.
Three cases are considered:  uniform lattice with periodic and open boundary conditions, and lattice with a parabolic potential. We find that in the latter case,  interplay of the potential and driving leads
to appearence of the  effective next-nearest neighbour couplings. In the uniform case with periodic boundary conditions the second- and third-order corrections to the averaged Hamiltonian are completely absent, while in the case with open boundary conditions they have a very simple form, found before in some particular cases  by S.Longhi  [Phys. Rev. B 77, 195326 (2008)].
These general results may found applications in designing effective Hamiltonian models  in experiments with ultracold atoms in optical lattices, e.g. for simulating solid-state phenomena.
\end{abstract}

\maketitle

\section{Introduction}

Effective Hamiltonians created by high-frequency perturbations have many interesting applications in physics; a well-known counterintuitive example of induced effective potential is provided by Kapitza pendulum \cite{Bogolyubov, Kapitza}. We are interested in applying  averaging methods of classical Hamiltonian mechanics  (see, e.g., \cite{AKN}) to quantum tight-binding models, which  often arise in solid-state and, more generally,  condensed-matter physics. 

In solid-state physics,  unusual transport phenomena may arise when an ac electric field is applied to the system, e.g. coherent desctruction of tunnelling and dynamic localization  \cite{Holthaus,Kenkre}. Corresponding applications to coherent control of tunnelling and electronic transport in semiconductor superlattices and arrays of coupled quantum dots have been receiving a lot of interest lately \cite{Korsch, Hohe}. Very recently, in many experiments with atoms in optical lattices,
effective Hamiltonians were created using high-frequency perturbations \cite{Holthaus,qtransport1, effective}.
A particle in a deep optical lattice potential can be described by a tight-binding model. Applying a high-frequency force,  one can engineer effective tunnelling constants in the model,
which can be useful to mimic certain solid-state phenomena \cite{effective}.  
For many realistic applications of such type, it is important to derive accurate effective Hamiltonians taking into account higher-order terms \cite{Longhi}. Here we find a useful method for such derivation in the spirit of canonical perturbation theory, and apply it for several tight-binding systems. Our approach is based on idea of canonical transformations removing time-dependence from the Hamiltonian, which in the present context means unitary transformations of square matrices.  Similar ideas have been applied to transport in classical periodic potentials \cite{transport1,transport2,transport3}.
In the next Section, the general method is outlined. In Section III,  it is applied to three different tight-binding models.Our approach is actually not limited to tight-binding systems, but it becomes especially transparent and elegant for such kind of systems.   Section IV gives concluding remarks.

\section{The averaging method}

Consider  a tight-binding model with the Hamiltonian

\be H  =  J \sum ( | n \rangle \langle n+1|    + | n+1 \rangle \langle n | ) + \sum \limits_n V(n) |n \rangle \langle n |  +  ed E(\omega t) \sum \limits_n n | n \rangle \langle n |,   \ee

where $J$ is the hopping parameter, $V(n)$ is the external potential (we consider only parabolic potential in this paper, $V(n) = V \frac{n^2}{2} $), $d$ is the intersite distance, $E$ is the applied electric field, $e$ is the charge of the particle.
The same model can be realized also with neutral particles, by approapriate shaking of the lattice.

Expanding a quantum state as $| \psi (t) \rangle = \sum c_n | n\rangle,$ one gets a system of equations

\be i \dot c_n = J (c_{n+1} +  c_{n-1} )  + V(n) c_n + {\cal E}(\omega t) n c_n \ee

It is convenient to  make a trasformation $c_n(t) = x_n(t) \exp \Bigl[ - i n \int\limits^t_0 {\cal E}(t') dt'  \Bigr]$, so that equations of motion are

\be i \dot x_n = J (x_{n+1} F(t) +  x_{n-1} F^*(t)) + V(n) x_n,   \ee
where $F(t) = \exp [ - i \int\limits_0^t { \cal E}(t') dt'  ] \  =  F_0 + \sum F_l  \exp(-i l \omega t),$   $ {\cal E} = edE$.

Introducing fast time $t'  = \omega t  \equiv t/\eps $, we get,  in the matrix form,

\be   i \dot X = \eps H X,  \ee  where

 \be H = J \begin{pmatrix}
   0 & F & \cdots & 0 \\
  F^* & 0 & \ddots & \vdots  \\
  \vdots  & \ddots  & \ddots & F \\
  0 & \cdots & F^* & 0
 \end{pmatrix}  \equiv  J( F {\cal U}  + F^* {\cal B}) ,  \label{open} \ee
where  $ {\cal U} , {\cal B}$ are  matrices  with unities on the first upper- and lower- codiagonals, correspodingly ($ {\cal U}_{mn}=\delta_{m,n+1} , {\cal B}_{mn}=\delta_{m,n-1}$).
 
Secondly, consider the case of the chain with periodic boundary conditions,  with  the Hamiltonian

\be H_p = J \begin{pmatrix}
   0 & F & 0 & \cdots & F^* \\
  F^* & 0 & \ddots  & \ddots & \vdots  \\
  \vdots  & \ddots  & \ddots &\ddots & 0 \\
   0 & 0 & \ddots & 0 & F\\
  F & 0 & \cdots & F^* & 0
 \end{pmatrix}  \label{periodic} \ee
 
 Thirdly, in the case of  a lattice with  additional parabolic potential ($V(n) =V \frac{n^2}{2}$) often employed in applications with ultracold atoms,
 the Hamiltonian is 
 
\be H_{pp} = J \begin{pmatrix}
   \frac{N^2 V}{2J} & F & 0 & \cdots & 0 \\
  F^* &  \frac{(N-1)^2 V}{2J}  & \ddots  & \ddots & \vdots  \\
  \vdots  & \ddots  & \ddots &\ddots & 0 \\
   0 & 0 & \ddots &  \frac{(N-1)^2 V}{2J}  & F\\
  0 & 0 & \cdots & F^* & \frac{N^2 V}{2J} 
 \end{pmatrix},  \label{parabolic} \ee where  $V$ is the strength of the parabolic potential, and the lattice has (2N+1) sites.
  
 In the spirit of the Hamiltonian averaging method in classical mechanics, we are making a  unitary transformation $X = C \tilde X$ so that equations for the transformed variables are
 
 \be i  \dot{ \tilde X} = [  C^{-1} \eps H C - i C^{-1} \dot C  ]   \tilde X . \ee
 
 We are looking for a transformation of the form $C = \exp[ \eps K_1 + \eps^2 K_2 + \eps^3 K_3 ], $
 where 
 $K_i$ are skew-Hermitian time-periodic  matrices, which would remove time-dependent terms from the Hamiltonian, leaving only time-independent terms.
 
 Generally, we have 
 
 \bea  C &\approx &I + \eps K_1 + \eps^2 \left( \frac{1}{2} K_1^2 + K_2 \right) + \eps^3 \left( \frac{1}{6} K_1^3 +\frac{1}{2}(K_1 K_2 + K_2 K_1) + K_3 \right) , \nonumber\\
       C^\dagger &\approx& I - \eps K_1  + \eps^2 \left( \frac{1}{2} K_1^2 - K_2 \right) + \eps^3 \left( -\frac{1}{6} K_1^3 +\frac{1}{2}(K_1 K_2 + K_2 K_1) - K_3 \right),  \eea
 where $I$ is the unity matrix.
 
 In the first order,  we have
 
 \be  i  \dot{  K_1 }  =  H(t)  -  \langle H(t) \rangle  \equiv  \left\{ H \right\}, \ee
 
 and therefore  $ i K_1 = \int  (H - \langle H \rangle ) dt  =  \int  \left\{ H \right\} dt . $
We introduce here curly brackets as taking time-periodic part of a time-dependent function: $\left\{ X \right\} \equiv  X -  \langle X(t) \rangle, $
where $\langle  X(t)   \rangle \equiv \frac{1}{2 \pi} \int \limits_0^{2\pi} X(t')  dt'$.

In the second order,
 
 \be  i  \dot{  K_2 }  = \left\{ H K_1 - K_1 H - \frac{i}{2} ( \dot{ K_1} K_1 - K_1 \dot{ K_1} )  \right\}.     \ee
 
In the third order,  we  finally get

\be  \eps H_{eff} = \eps H_1 + \eps^2 H_2 + \eps^3 H_3 ,\ee

where

\bea
H_1 &=&   \langle H \rangle \nonumber\\
H_2 &=& \langle H K_1 - K_1 H - \frac{i}{2} (\dot  K_1 K_1 - K_1 \dot  K_1)  \rangle  \label{formula} \\
H_3 & =& \langle H K_2 - K_2 H + \frac{1}{2} (H K_1^2 + K_1^2 H) - K_1 H K_1  - \frac{i}{2}(\dot  K_1 K_2 - K_1 \dot  K_2 + \dot  K_2 K_1 - K_2 \dot  K_1  ) \nonumber\\
  &-& \frac{i}{6} (\dot K_1 K_1^2 + K_1^2 \dot K_1 -
2 K_1 \dot K_1 K_1)     \rangle  \nonumber
\eea

These general formulas can be applied to particular models, as done in the next Section.

One can also write expressions  Eq.(\ref{formula}) in a more compact way:

\bea
H_1 &=&   \langle H \rangle \nonumber\\
H_2 &=&  \frac{1}{2} \langle [ \{ H \} ,K_1]  \rangle   \label{formula2} \\
H_3 & =& \langle [H, K_2] + \frac{1}{2}[ [H,K_1],K_1 ]  - \frac{i}{2}( [\dot  K_1, K_2 ] + [ \dot  K_2, K_1 ]  )  - \frac{i}{6} [ [ \dot{K_1},K_1],K_1]  \rangle, \nonumber
\eea
where square brackets denote matrix commutation: $[A,B] = AB -BA$.

\section{Applications to particular models}
For the uniform model with  periodic boundary conditions (\ref{periodic}), we get a very interesting and important result: $H_2 = H_3 = 0.$
First- and second-order corrections are completely absent in this case (note that, since the Hamiltonian $\eps H_1$ contains $\eps$, $H_2$ and $H_3$ define
 the first and the second-order corrections, correspondingly).

For the uniform model with open boundary conditions (\ref{open}),  we have

$ K_1 = - i \int  \left\{ H \right\} dt =  J \begin{pmatrix}
   0 & L & \cdots & 0 \\
  -L^* & 0 & \ddots & \vdots  \\
  \vdots  & \ddots  & \ddots &L \\
  0 & \cdots & -L^*& 0
 \end{pmatrix} = J( L {\cal U} - L^* {\cal B}), \quad  L  \equiv \sum \limits_{l \ne 0}  \frac{F_l}{l} \exp(-ilt) $

$ \dot K_1 = - i   \left\{ H \right\}  = -i J \begin{pmatrix}
   0 & \tilde F & \cdots & 0 \\
  \tilde F^* & 0 & \ddots & \vdots  \\
  \vdots  & \ddots  & \ddots &\tilde F \\
  0 & \cdots & \tilde F^*& 0
 \end{pmatrix} = - i J( \tilde F {\cal U}  + \tilde F^* {\cal B}), \quad  \tilde F \equiv \left\{ F \right\}  = \sum \limits_{l \ne 0}  F_l \exp(-ilt)$
 
\be H K_1 - K_1 H = J^2  \begin{pmatrix}
   -P & 0 & \cdots & 0 \\
  0 & 0 & \ddots & \vdots  \\
  \vdots  & \ddots  & \ddots & 0  \\
  0 & \cdots & 0  & P
 \end{pmatrix} \equiv -J^2 P {\cal Z}_1, \quad  P  \equiv L F^* + F L^*,  \ee
 where $ {\cal Z}_1$ is a square matrix with $1,-1$ in  the upper left  and the bottom right corners, and zeros elsewhere.

\be \dot K_1 K_1 - K_1 \dot K_1 =  i J^2 D  {\cal Z}_1 \quad  D  \equiv  L \tilde F^* + \tilde F L^* \ee

\be K_2 = - i J^2  {\cal Z}_1 T, \quad  T  \equiv \int \left\{ -P + \frac{D}{2} \right\} dt \equiv   \int S dt, \ee

 \be \dot K_2 = - i J^2  {\cal Z}_1 S  \ee
 
 \be H K_2 - K_2 H  = i T J^3  \begin{pmatrix}
   0 & F & 0 & \cdots & 0 \\
  -F^* & 0 & \ddots  & \ddots & \vdots  \\
  \vdots  & \ddots  & \ddots &\ddots & 0 \\
   0 & 0 & \ddots & 0 & F\\
  0 & 0 & \cdots &- F^* & 0
 \end{pmatrix}   =  i T J^3 (F  {\cal U}_1 - F^* {\cal B}_1 ), \ee
 where ${\cal U}_1,{\cal B}_1$ are  matrices with only two non-zero entries '1'  on the ends of the upper- and lower-  co-diagonal.

\be \frac{1}{2} (H K_1^2 + K_1^2 H ) - K_1 H K_1 = - \frac{P J^3}{2} (L {\cal U}_1 + L^* {\cal B}_1) \ee

\be \frac{i}{6} (\dot K_1 K_1^2 + K_1^2  \dot K_1 - 2 K_1 \dot K_1 K_1 )  = - \frac{D J^3}{6} (L {\cal U}_1 + L^* {\cal B}_1) \ee

\be \frac{i}{2} (\dot K_1 K_2 - K_1  \dot K_2  +  \dot K_2  K_1 - K_2 \dot K_1 )  =  \frac{ J^3}{2}  (S[L {\cal U}_1 + L^* {\cal B}_1] +i T [\tilde F  {\cal U}_1 - \tilde F^* {\cal B}_1 ]). \ee

One obtains

\be
\langle P \rangle = \langle D \rangle =  2 \sum \limits_{l=1} \frac{|F_l|^2 - |F_{-l}|^2}{l}  \equiv 2 D_2.
\ee

\be
\langle LD \rangle = \sum \limits_{k \ne 0} \sum \limits_{l \ne 0} \left(   \frac{F_k F_l F^*_{k+l} + F_k F_l^* F_{l-k}}{kl} \right) \equiv L_3,
\ee

\be
\langle T F \rangle =  \frac{i}{2} L_3.
\ee

To conclude,  in the case of  open boundary conditions effective Hamiltonians have a very simple form

\be H_2 = J^2  D_2  {\cal Z}_1,  \quad H_3 = -\frac{J^3}{3}  (L_3 {\cal U}_1 + L_3^* {\cal B}_1). \ee

Thirdly,  in the model with parabolic potential (\ref{parabolic}) we have

\be
K_1 =  J( L {\cal U} - L^* {\cal B}), \quad  \dot K_1 = J( \tilde F {\cal U}  + \tilde F^* {\cal B}),
\ee

\be \dot K_1 K_1 - K_1 \dot K_1 =  i J^2 D  {\cal Z}_1  \ee

\be
H K_1 - K_1 H 
= J^2  \begin{pmatrix}
   -P &  \frac{2N-1}{2} \frac{V}{J}  L & 0 & \cdots & 0 \\
   \frac{2N-1}{2} \frac{V}{J}  L^*& 0 & \frac{2N-3}{2} \frac{V}{J}  L& 0 & \vdots  \\
  \vdots  & \ddots  & \ddots &\ddots & 0 \\
   0 & 0 & \ddots & 0 &  -\frac{2N-1}{2} \frac{V}{J}  L\\
  0 & 0 & \cdots & - \frac{2N-1}{2} \frac{V}{J}  L^* & P
 \end{pmatrix}  \ee

The first correction to the averaged Hamiltonian $\eps^2 H_2$  looks exactly the same as that of the uniform case, and does not depend on the potential.
In the following, we neglect  influence of the boundary conditions, assuming the lattice is very long.
Then, the first correction is absent,  while the second correction $\eps^3 H_3 = \eps^3 J^2 V M$ contains contribution from the parabolic potential. $M$ is a 5-diagonal matrix,  with the following  entries (non-zero diagonals are listed from top to  bottom,  with $('0')$  denoting the main diagonal,  $('+2')$ and $('+1')$ upper co-diagonals,  $('-2')$ and $('-1')$
 lower co-diagonals) :

\bea
&('+2')&  (  i   \langle  F L_2  \rangle    - \frac{i}{2}    \langle \tilde F L_2  \rangle ) \delta_{m,n+2}  \nonumber\\
&('+1')&                            0                 \nonumber\\
&('0')& 		(-  i   \langle  F^* L_2 + F L_2^* \rangle   	+   \frac{i}{2}  \langle \tilde F^* L_2 + \tilde F L_2^*  \rangle	)\delta_{m,n}  	 \\
&('-1')& 			0		   \nonumber\\
&('-2')&	(  i   \langle  F^* L^*_2  \rangle   	 - \frac{i}{2}    \langle  \tilde F^* L^*_2  \rangle )  	\delta_{m,n-2},		   \nonumber
\eea
$L_2 = i \sum \limits_{l \ne 0} \frac{F_l}{l^2} \exp(-ilt)$, $  \langle  L_2 \rangle=0$, 
  $   \langle  F L_2  \rangle =   \langle  \tilde F L_2  \rangle =  i \sum \limits_{l \ne 0} \frac{F_l F_{-l}}{l^2}$
 
One can see that this correction  creates next-nearest-neighbour couplings: non-zero entries are not on the main co-diagonals, as it would be in case of nearest-neighbour couplings,
but on the next-to main co-diagonals.
Since $   \langle  F^* L_2 + F L^*_2  \rangle =   \langle \left(\sum \limits_{l \ne 0} F^*_l \exp(i lt) i \sum \limits_{m \ne 0} \frac{F_m}{m^2} \exp(-imt) \right) +  \left( \sum \limits_{l \ne 0} F_l \exp(-i lt)( -i) \sum \limits_{m \ne 0} \frac{F^*_m}{m^2} \exp(imt) \right)  \rangle =0$,
finally  the second correction has a very simple, two-diagonal form

\bea
\eps^3 H_3^{mn} &=& \eps^3 J^2 V \frac{i}{2} ( \langle  F L_2 \rangle  \delta_{m,n+2}  + \langle  F^* L^*_2 \rangle  \delta_{m,n-2}    )   \\
     &=& - \frac{\eps^3}{2} J^2 V \left(  \sum_{l \ne 0}  \delta_ {m,n+2}  \frac{F_l F_{-l}}{l^2}    +  \delta_{m,n-2} \frac{F_l^* F_{-l}^*}{l^2}   \right)
\eea

Consider a particular case of harmonic driving,  with ${ \cal E} = {\cal E}_0 \cos t $.  
We have $F_l = J_l ({\cal E}_0 )$.
The induced next-nearest neighbour coupling is $J' = - \eps^3  J^2 V \sum \limits_{l>0} \frac{(-1)^l J_l^2({\cal E}_0 )}{l^2} $ 

Returning from the fast time back to the original time, we have

\be J' = -   \frac{J^2 V}{\omega^2} \sum \limits_{l>0} \frac{(-1)^l J_l^2({\cal E}_0 )}{l^2} 
\ee

As a function  of  ${\cal E}_0$, it has an oscillatory form, and one can choose parameters that nullify the next-neighbour coupling (e.g.,  ${\cal E}_0 = 3.32, 4.11,$ etc), or
maximize it  (e.g.,  ${\cal E}_0 = 1.77, 5.24,$ etc).  It can be tuned to be either positive or negative, which  may be useful for applications. 

\section{Conclusions}

The approach based on canonical transformations and described in Section II has been applied to three different lattice  systems:  uniform lattice with open boundary conditions, uniform lattice with periodic boundary conditions, and a lattice with 
an additional parabolic potential.  In the first case, we generalize results obtained by S. Longhi \cite{Longhi}. In particular, we show that second-order corrections have very simple ('boundary') form.
In the second case, we get a very interesting and unexpected result: absence of corrections to the averaged Hamiltonian in the second and third order.
In the case of external parabolic potential,  another unexpected result is found:   interplay of driving and non-uniform external potential creates effective (uniform!) next-nearest neighbour couplings.  The same result can be obtained in the semiclassical approach  \cite{IMathey}.
These results, we believe, may found applications in forthcoming experiments with cold atoms in driven optical lattices.
 
This work was partially supported by RFBR (project no. 13-01-00251).  A.P.I    thanks  A.Polkovnikov, M.Thorwart, A.Eckardt, A.Engel and  L.Mathey for interesting and simulating discussions.


\begin{thebibliography}{10}
\bibitem{Bogolyubov} 
N.N. Bogolyubov, The Theory of Perturbations in Nonlinear Mehanics, Proc. Inst. Struct. Mech., no. 14,  9 (1950)
 [In Russian: Teoriya vozmusheniy v nelineynoy mekhanike. Sbornik trudov Instituta
stroitel'noy mekhaniki AN USSR, no. 14, 9 (1950) ].

\bibitem{Kapitza} P.L. Kapitza, Dynamic stability of a pendulum when its point
of suspension vibrates,  Soviet Phys. JETP 21, 588 (1951).


\bibitem{AKN}
V. I. Arnold, V. V. Kozlov, and A. I. Neishtadt, Mathematical Aspects of Classical and Celestial Mechanics, 3rd ed. (Springer, Berlin, 2006).

\bibitem{Kenkre} D.H.Dunlap and V.M. Kenkre, Phys. Rev. B  {\bf 34}, 3625 (1986).
\bibitem{Holthaus} M. Holthaus, Phys. Rev. Lett. {\bf 69}, 351 (1992).
\bibitem{Hohe} M. Holthaus, D.W. Hohe, Philos. Mag. B {\bf 74}, 105 (1996).
\bibitem{Korsch} M. GlŸck, A.R. Kolovsky, and  H.J. Korsch, Phys. Rep. {\bf  366 }, 103 (2002).

\bibitem{effective}
N. Strohmaier, Y. Takasu, K. Gunter, R. Jordens, M. Kohl, H. Moritz, and T. Esslinger, Phys. Rev. Lett. 99, 220601 (2007); A. Alberti et al., Nat. Phys. 5, 547 (2009); A. Zenesini et al., Phys. Rev. Lett. 102, 100403 (2009); A. Eckardt et al., Europhys. Lett. 89, 10010 (2010).
\bibitem{qtransport1} K. Kudo, T. Boness, and T. S. Monteiro, Phys. Rev. A 80, 063409 (2009); K. Kudo and T. S. Monteiro, ibid. 83, 053627 (2011); A. R. Kolovsky, E. A. Gomez, and H. J. Korsch, ibid. 81, 025603 (2010).

\bibitem{Longhi} S. Longhi, Phys. Rev. B 77, 195326 (2008)


\bibitem{transport1}
A. P. Itin, R. de la Llave, A. I. Neishtadt, and A. A. Vasiliev, Chaos 12, 1043 (2002).
\bibitem{transport2}
X. Leoncini, A. I. Neishtadt, and A. A. Vasiliev, Phys. Rev. E 79, 026213 (2009).
\bibitem{transport3}
A. P. Itin and A. I. Neishtadt, Phys. Rev. E 86, 016206 (2012)


\bibitem{IMathey} A.P.Itin,  L.Mathey,  in preparation.
 

\end{thebibliography}
\end{document}